\documentclass[twoside]{dis08}
\usepackage[latin1]{inputenc}
\usepackage[dvips]{graphicx,epsfig,color}
\usepackage{wrapfig,rotating}
\usepackage{amssymb,amsmath,array}

\def\cascade{{\sc Cascade}}
\def\herwig{{\sc Herwig}}
\def\disent{{\sc Disent}}
\def\pythia{{\sc Pythia}}

\begin{document}
\title{
Dijet azimuthal distributions and initial-state parton showers}

\author{F.~Hautmann$^1$ and H.~Jung$^2$
%
%
\vspace{.3cm}\\
%
1 - Oxford University, Theoretical Physics Department \\
Oxford OX1 3NP, UK 
%
\vspace{.1cm}\\
2 - Deutsches Elektronen Synchrotron \\
Hamburg D-22603, Germany\\
}

\maketitle

\begin{abstract}
 We investigate angular correlations in multi-jet final states 
 at high-energy colliders  and discuss their sensitivity to initial-state 
showering effects, including   QCD  coherence and corrections to 
 collinear ordering~\cite{url}. 
\end{abstract}

\vskip 0.5 cm 

\hskip 0.8 cm {\em   Presented at the Workshop DIS08, 
University College London,  April 2008}

\vskip 0.5 cm 

\begin{figure}
\centerline{\includegraphics[width=0.35\columnwidth]{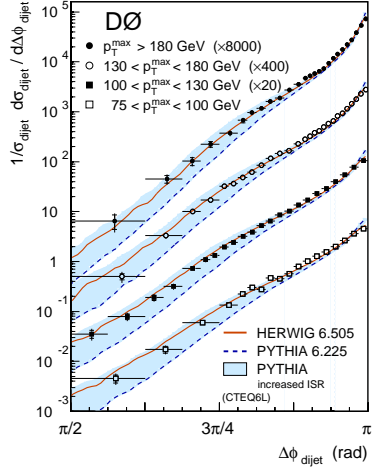}}
\caption{\label{Fig:d0az} Dijet azimuthal  correlations measured by 
D0 along with the \herwig\  and  \pythia\  results~\protect\cite{d02005}.}
\end{figure}

Events with multiple hadronic  jets are central to many 
aspects of the LHC physics program and their analysis 
will require  realistic Monte Carlo simulations. See e.g.~\cite{alwalletal}.  
In a multi-jet event  the   correlation in the azimuthal angle   $\Delta \phi$,   
defined to be   between the two hardest jets,     provides  a   useful  measurement,   
sensitive to how well QCD   multiple-radiation  effects are described,  
 and has been  used to tune shower Monte Carlo event generators~\cite{albrow}.   
     The  Tevatron $\Delta \phi$  measurements~\cite{d02005}  
  admit a reasonable description by 
  Monte Carlo, see        \herwig\      and 
  \pythia\               results 
  in Fig.~\ref{Fig:d0az}~\cite{d02005}.   In particular   the data 
    are  sensitive to 
     initial-state showering parameters and have been used 
     for re-tuning of these parameters 
     in \pythia~\cite{albrow}. On the other hand,    the 
     {\small HERA}  $\Delta \phi$  measurements~\cite{h1deltaphi,zeus1931}  
  are not   well described by  the standard 
   \herwig\    and   \pythia\      Monte Carlo showers    in  most 
 of the data kinematic    range       (see below). 

  At the LHC, measurements of   $\Delta \phi$ distributions in multi-jet  events 
  may  become  accessible    relatively early.  
  Such  complex  hadronic final states at  LHC energies are potentially sensitive to 
  corrections  to the collinear ordering 
    implemented in standard 
parton showers~\cite{ictppaper}. 
      In particular,  for jets of  given $E_T$ the   partonic  momentum  fraction 
      $x$ is reduced 
as the energy increases, and angular   correlations 
probe   coherence  effects in the     spacelike  branching~\cite{hj_angjet},  associated with 
non-collinear radiation at  $ x \ll 1$  and  not  included in    \herwig\    or   \pythia.

      Monte Carlo generators  designed to take  these effects   into account 
 are based (see e.g.~\cite{jepp06,hann04} and early studies in~\cite{mw}) on 
 transverse-momentum dependent   parton distributions and 
 matrix elements, defined via high-energy factorization~\cite{hef}.      
  General formulations for these distributions in initial-state showers 
  are studied in~\cite{collinszu}. 

\begin{figure}[h!]
\centerline{\includegraphics[width=0.45\columnwidth]{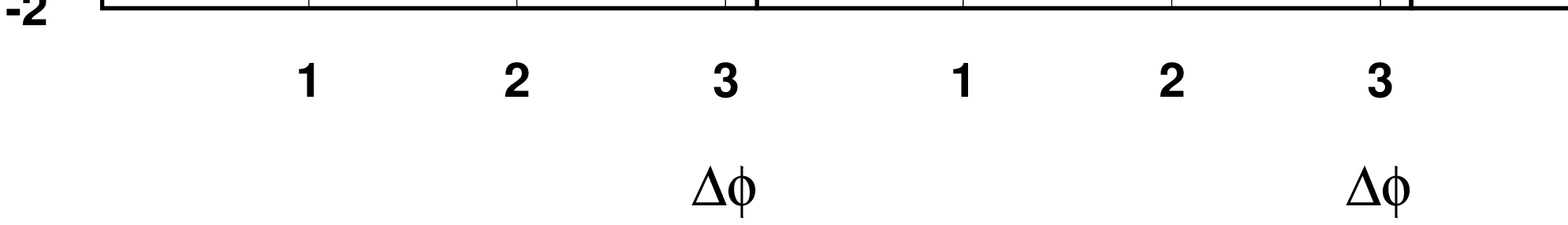}
	    \includegraphics[width=0.45\columnwidth]{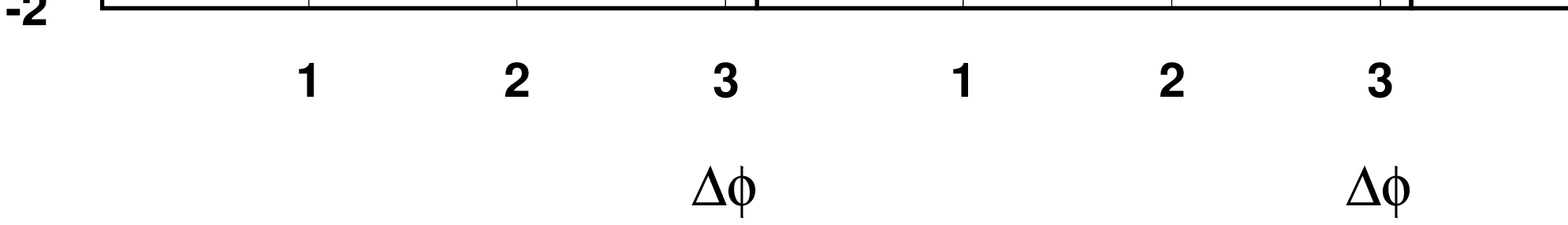}}
\vspace*{-5.5 cm}
\caption{\label{Fig:azz} Azimuthal  correlations~\protect\cite{hj_angjet}   by the 
k$_\perp$-shower 
\cascade\ and by \herwig\ compared 
with the ep data~\protect\cite{zeus1931}: 
(left) two-jet cross section; (right) three-jet cross section.}
\end{figure}

\begin{figure}[h!]
\centerline{\includegraphics[width=0.45\columnwidth]{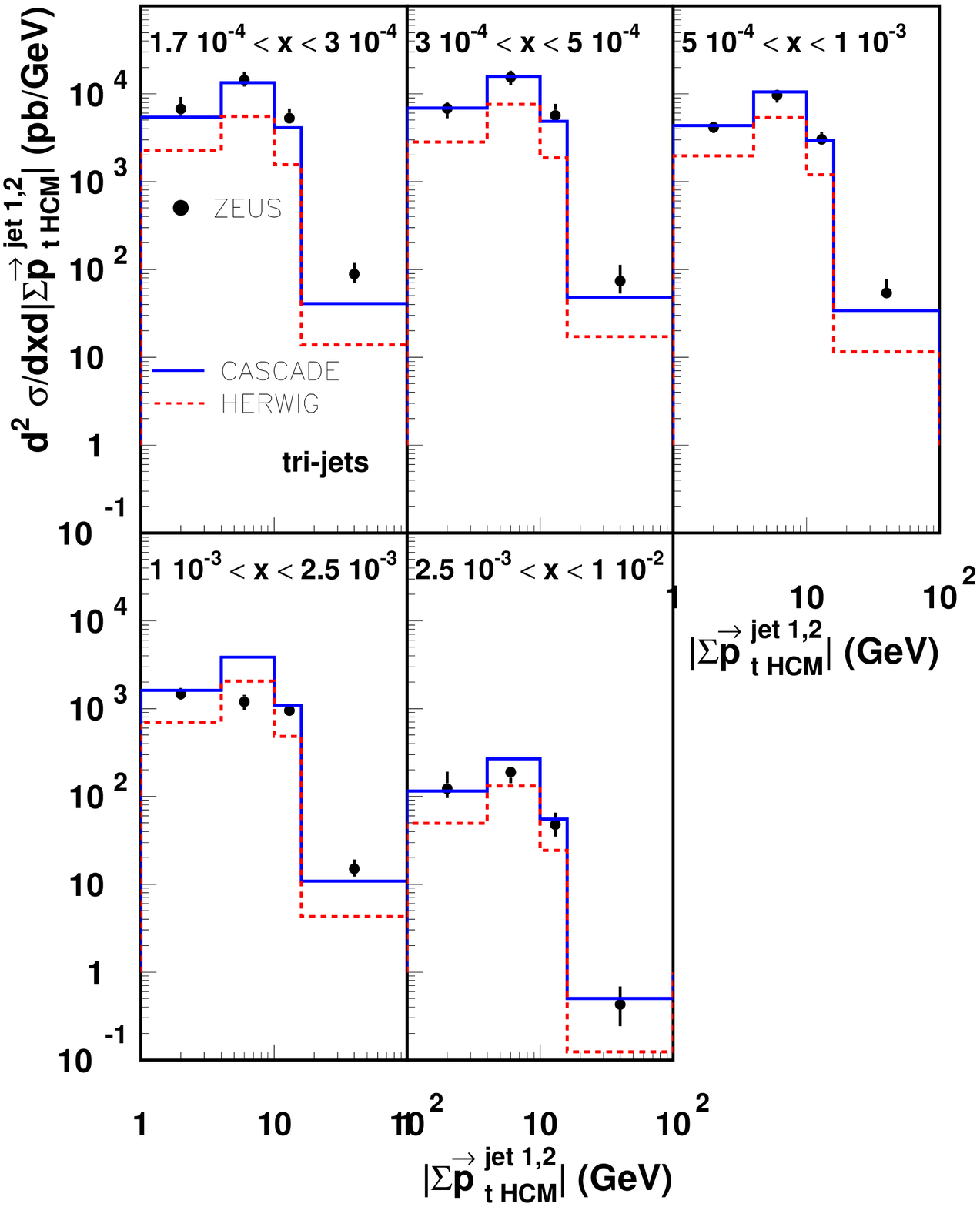}
	    \includegraphics[width=0.45\columnwidth]{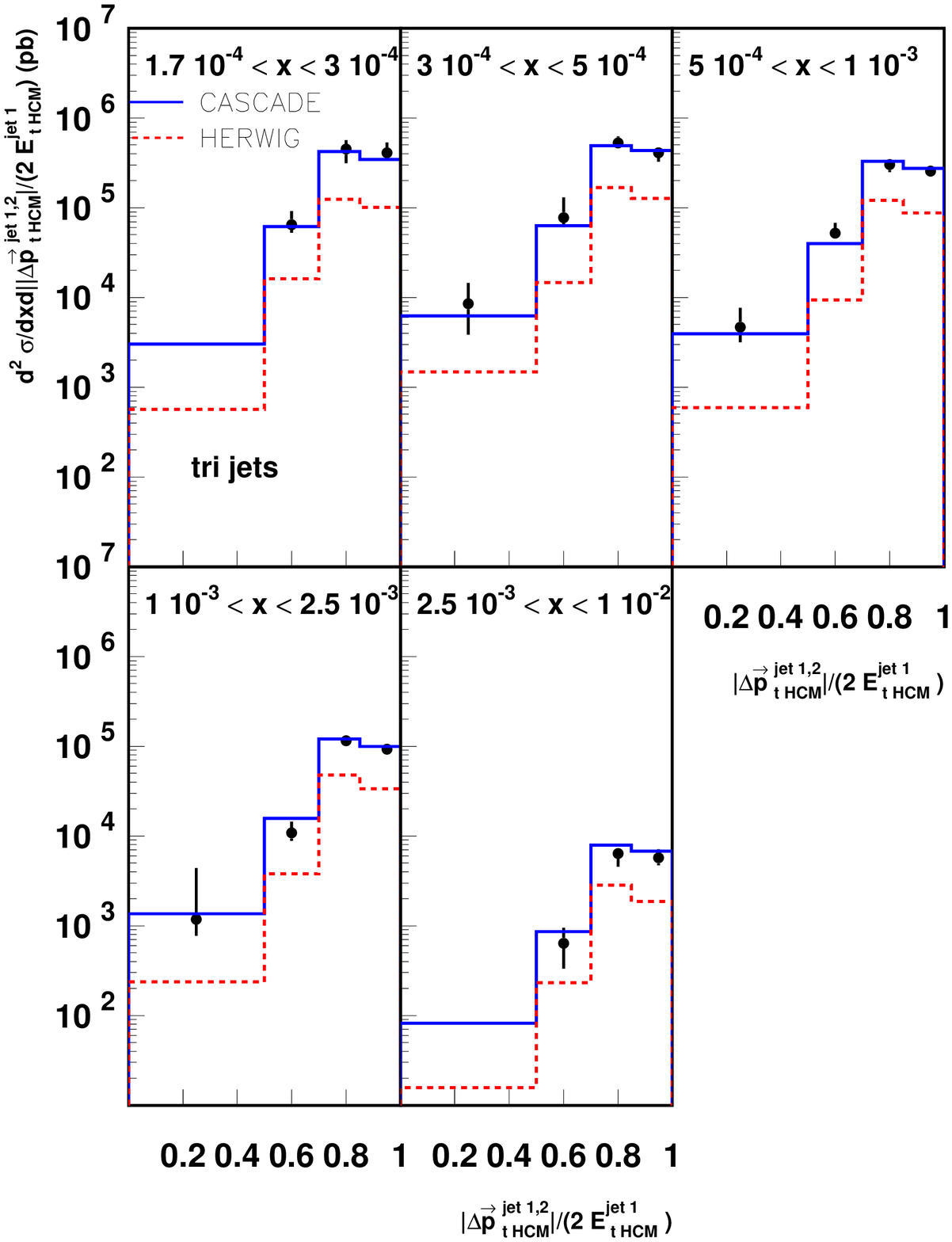}}
\caption{\label{Fig: 3} Transverse momentum correlations \protect\cite{hj_angjet} 
by the k$_\perp$-shower \cascade\ and by \herwig\  compared 
with the 3-jet data~\protect\cite{zeus1931}. The variables $\Sigma p_t$ (left) and 
$\Delta p_t$  (right)  are  as defined in~\protect\cite{zeus1931,hj_angjet}.}
\end{figure}

Ref.~\cite{hj_angjet} investigates the effects of corrections to 
collinear-ordered showers 
on    correlations in multi-jet final  states, using the precise 
ep measurements~\cite{zeus1931} that have recently become available.  These   
measurements 
 are characterized by    large phase space available for jet 
production and  by small $x$ kinematics,  potentially 
 relevant for extrapolation of 
initial-state showering effects to the LHC. In 
Fig.~\ref{Fig:azz} we report  results~\cite{hj_angjet} for  the azimuthal 
$\Delta \phi$ distribution 
in two-jet and three-jet cross sections.  
In Fig.~\ref{Fig: 3} we give  
results   for  the   $\Sigma p_t$  and $\Delta p_t$ 
distributions~\cite{zeus1931,hj_angjet}  measuring 
   the transverse-momentum imbalance between 
the leading jets.

These results  show   that 
the shape of the distributions is different for \herwig\ and for the 
k$_\perp$-shower Monte Carlo  \cascade~\cite{jung02}, with the largest differences 
occurring at small $\Delta \phi$ and small $\Delta p_t$, where the   two highest $E_T$ 
jets are far from back to back 
and one has  effectively  three hard,  well-separated jets.
Ref.~\cite{hj_angjet} also analyzes the  angular  distribution of the third jet 
and finds significant contributions from 
 regions where the transverse momenta in the initial state 
shower are not ordered.  The description of the  
measurement by the k$_\perp$-shower is 
 good, whereas   the collinear-based  \herwig\  shower  is not
sufficient to describe it.  

\begin{figure}[h!]
\centerline{\includegraphics[width=0.45\columnwidth]{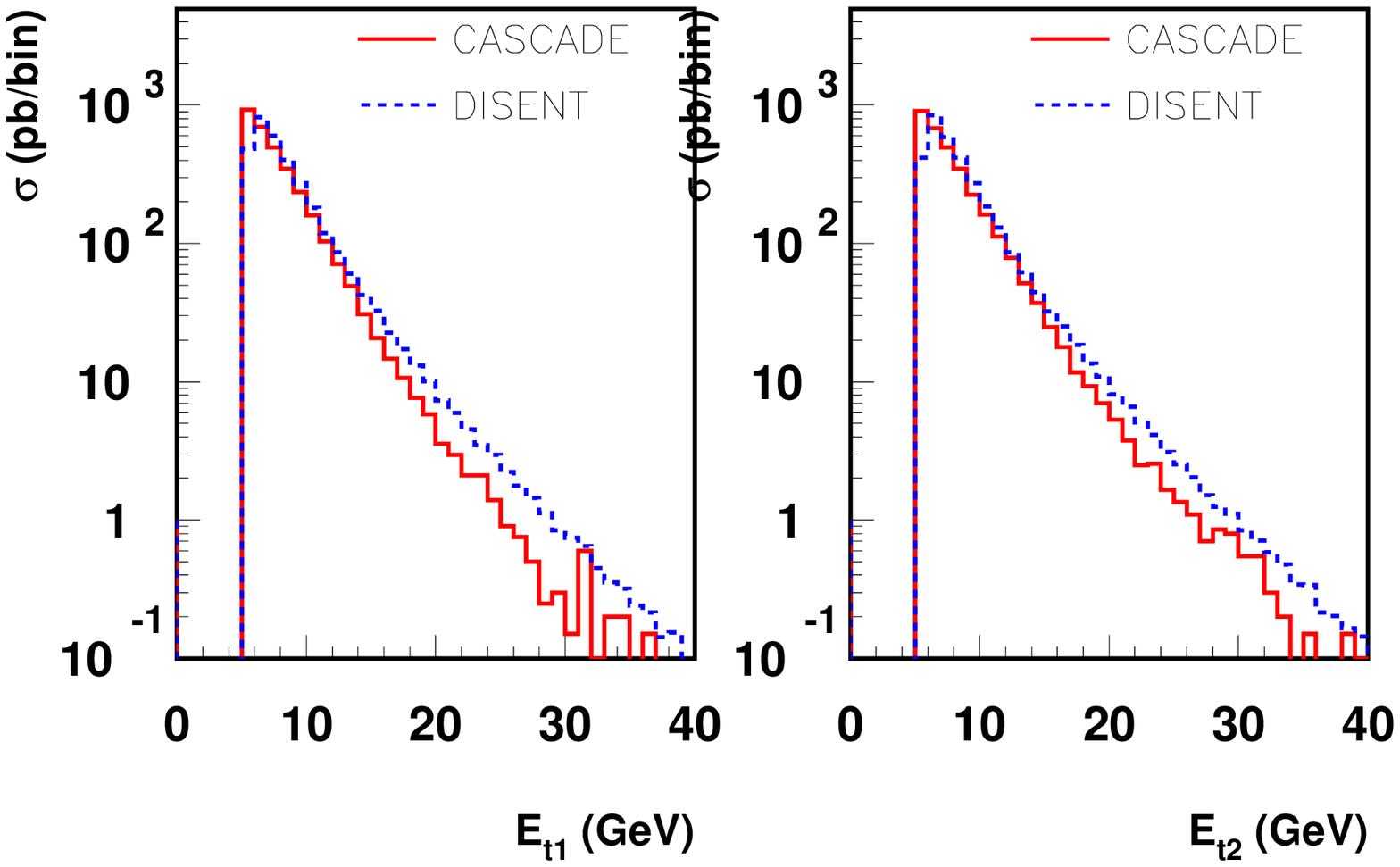}
	    \includegraphics[width=0.45\columnwidth]{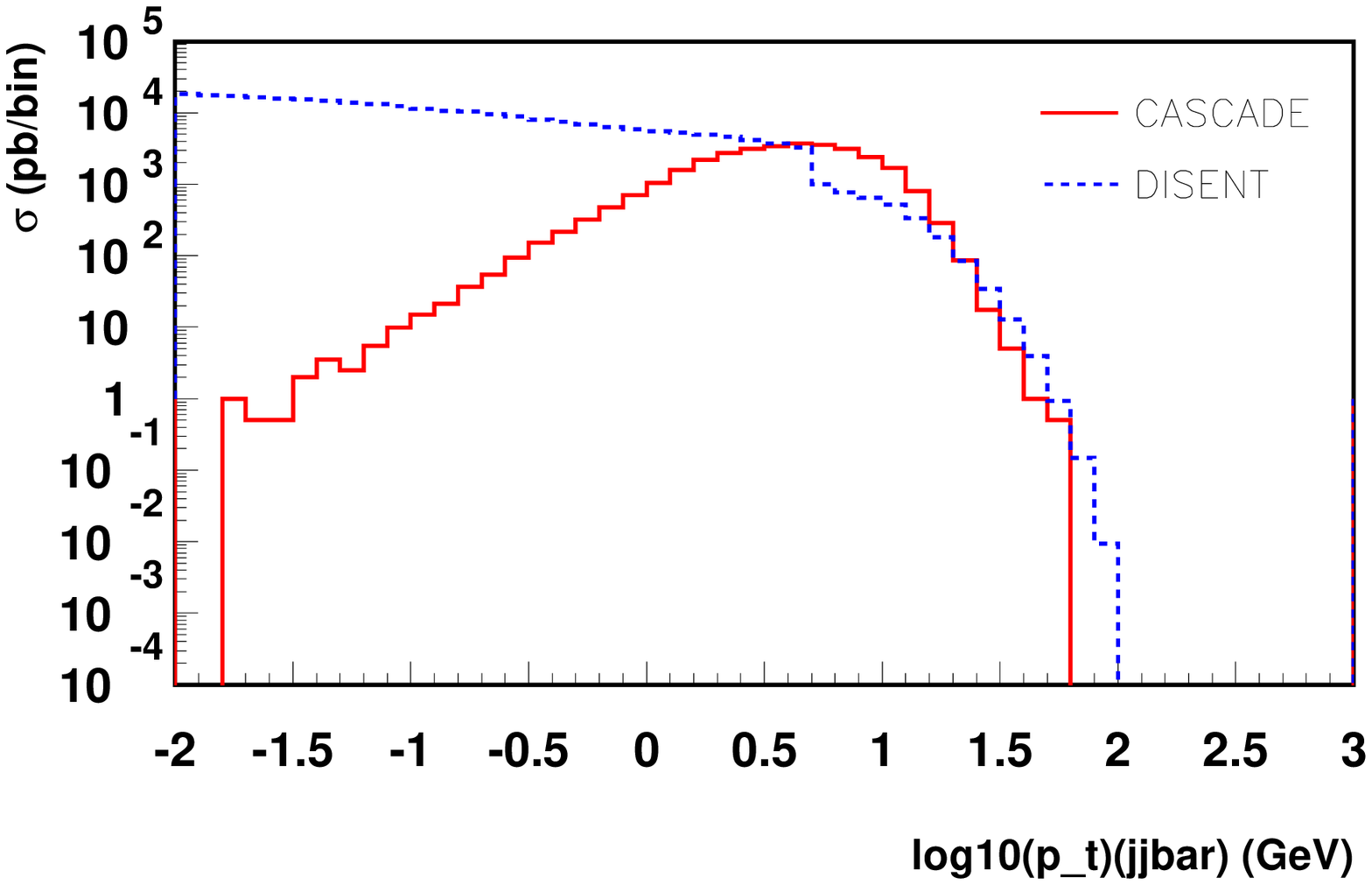}}
\vspace*{-3.5 cm}
\caption{\label{Fig: 4} Comparison of the 
k$_\perp$-shower   \cascade\   with the NLO  di-jet calculation \disent: 
(left)  single-jet distributions; (right) di-jet distributions.}
\end{figure}

The physical picture underlying 
the k$_\perp$-shower method involves both 
 transverse momentum dependent  pdfs and  matrix elements~\cite{ictppaper}. 
 The angular and momentum 
correlations of Figs.~\ref{Fig:azz},\ref{Fig: 3} are found~\cite{hj_angjet,hjradcor} to be 
 sensitive  in particular 
 to the large-k$_\perp$ tail   in  the hard matrix elements~\cite{hef}. 
More detailed studies of these   
off-shell contributions 
are currently underway, including comparisons with results of 
next-to-leading order  (NLO)
event generators,  see  single-jet and  di-jet distributions  in Fig.~\ref{Fig: 4}.  
Here we see in particular that the dijet   $p_t$ spectrum at  high  $p_t$ 
is close for the NLO calculation and the 
k$_\perp$-shower (at low $p_t$ we see the effect of the Sudakov form factor in 
the shower). 
Ref.~\cite{hj_angjet}  illustrates that the collinear approximation to the 
matrix element does not describe the shape of the angular distribution 
at small $\Delta \phi$. 
 We note  that the inclusion of  the  perturbatively computed  high-k$_\perp$ 
 correction    
 distinguishes the  calculation~\cite{hj_angjet}  of  multi-jet cross sections 
 from other  shower approaches (see e.g.~\cite{hoeche})  
 that include transverse momentum 
dependence in the  pdfs but not  in the  matrix elements.

It is worth emphasizing that 
the  coherence effects  in the angular distributions computed    above  
are associated with multi-gluon radiation terms to the  initial-state shower  that become 
non-negligible at 
  high energy (small $x$) and small $\Delta \phi$.    These   can be incorporated 
  using  the formulation at fixed transverse momentum.    
Near the  back-to-back  region  of large $\Delta \phi$~\cite{delenda}, 
 corrections due to 
  mixed Coulomb/radiative terms     
  may  also  become  important and affect the basic picture: see recent 
  studies in~\cite{0708pap}. 
 See also~\cite{manch} for  a  related discussion  of  Coulomb 
 contributions.   More general issues on  
unintegrated  pdfs   in parton showers  
  are discussed  in~\cite{ictppaper,collinszu,endp}. 
 Applications to semi-inclusive processes and 
 spin asymmetries are reviewed in~\cite{murgiarev}.

Besides jet final states,   
the corrections to collinear-ordered showers 
 discussed in this article will also be relevant to 
 heavy particle production~\cite{hann04,hef,hgs}, including 
 phenomenological studies of small-$x$ broadening in W and Z  
 $p_\perp$ distributions~\cite{cpyuan1},   kinematical relations of 
 DIS event shapes with Drell-Yan production~\cite{dasgqt}, heavy flavor 
 production. First  results on top-antitop pair production at the LHC may be found 
 in the first paper of reference~\cite{ictppaper}.


\begin{footnotesize}



%

\end{footnotesize}


\end{document}